\documentclass[
 reprint,
 amsmath,amssymb,
 nofootinbib,
 aps,
 prd,
 superscriptaddress,
 floatfix,
]{revtex4-1}
\usepackage{aas_macros}
\usepackage{graphicx}
\usepackage{dcolumn}
\usepackage{bm}
\usepackage{amsmath}
\usepackage{float}
\usepackage{lipsum}
\usepackage{xcolor}
\usepackage{times}
\usepackage{textcomp}
\usepackage{latexsym}
\usepackage[hidelinks]{hyperref}
\usepackage{mathtools}
\usepackage{url}
\usepackage{aas_macros}
\usepackage[utf8]{inputenc}
\usepackage[normalem]{ulem}
\usepackage{xspace}
 \usepackage{caption}
 \usepackage{booktabs}
 \usepackage{natbib,tabularx}
\usepackage{hyperref}

\definecolor{lightgreen}{RGB}{179,255,164}
\definecolor{lightblue}{RGB}{196,206,255}
\definecolor{light-gray}{RGB}{230,230,230}
\definecolor{carl-orange}{RGB}{217,95,2}

\newcommand{\vt}{\ensuremath{\boldsymbol{\theta}}\xspace}

\newcommand{\laa}{\ensuremath{\boldsymbol{\lambda}^A}\xspace}
\newcommand{\lac}{\ensuremath{\boldsymbol{\lambda}^C}\xspace}

\newcommand{\err}{\ensuremath{R(f)}\xspace}

\newcommand{\beq}{\begin{equation}}
\newcommand{\eeq}{\end{equation}}
\newcommand{\beqa}{\begin{eqnarray}}
\newcommand{\eeqa}{\end{eqnarray}}
\newcommand{\linf}{\texttt{LALInference}\xspace}

\newcommand{\dea}{\ensuremath{\delta A}\xspace}
\newcommand{\dep}{\ensuremath{\delta \phi}\xspace}
\newcommand{\deaI}{\ensuremath{\delta A^I}\xspace}
\newcommand{\depI}{\ensuremath{\delta \phi^I}\xspace}
\newcommand{\pz}{\textit{physiCal}\xspace}
\newcommand{\scal}{\textit{splineCal}\xspace}
\newcommand{\ud}{\ensuremath{\mathrm{d}}\xspace}

\newcommand{\hpg}{\ensuremath{\eta_{\mathrm{NIST}}}\xspace}

\newcommand{\LIGOlabMIT}{\affiliation{LIGO Laboratory, Massachusetts Institute of Technology, 185 Albany St, Cambridge, MA 02139, USA}}
\newcommand{\MKI}{\affiliation{Department of Physics and Kavli Institute for Astrophysics and Space Research, Massachusetts Institute of Technology, 77 Massachusetts Ave, Cambridge, MA 02139, USA}}
\newcommand{\LIGOlabCaltech}{\affiliation{LIGO Laboratory, California Institute of Technology, Pasadena, California 91125, USA}}

\begin{document}

\title{\pz: A physical approach to the marginalization of LIGO calibration uncertainties}

\author{Salvatore Vitale}
\email[]{salvo@mit.edu}
\LIGOlabMIT 
\MKI
\author{Carl-Johan Haster}
\email[]{haster@mit.edu}
\LIGOlabMIT 
\MKI
\author{Ling Sun}
\LIGOlabCaltech
\affiliation{OzGrav-ANU, Centre for Gravitational Astrophysics, College of Science, The Australian National University, ACT 2601, Australia}
\author{Ben Farr}
\affiliation{Department of Physics, University of Oregon, Eugene, OR 97403, USA}
\author{Evan Goetz}
\LIGOlabCaltech
\affiliation{University of British Columbia, Vancouver, BC V6T 1Z4, Canada}
\author{Jeff Kissel}
\affiliation{LIGO Hanford Observatory, Richland, WA 99352, USA}
\author{Craig Cahillane}
\LIGOlabCaltech

\date{\today}

\begin{abstract}
The data from ground based gravitational-wave detectors such as Advanced LIGO and Virgo must be calibrated to convert the digital output of photodetectors into a relative displacement of the test masses in the detectors, producing the quantity of interest for inference of astrophysical gravitational wave sources.
Both statistical uncertainties and systematic errors are associated with the calibration process, which would in turn affect the analysis of detected sources, if not accounted for. 
Currently, source characterization algorithms either entirely neglect the possibility of calibration uncertainties or account for them in a way that does not use knowledge of the calibration process itself. 
We present \pz, a new approach to account for calibration errors during the source characterization step, which directly uses all the information available about the instrument calibration process. 
Rather than modeling the overall detector's response function, we consider the individual components that contribute to the response. 
We implement this method and apply it to the compact binaries detected by LIGO and Virgo during the second observation run, as well as to simulated binary neutron stars for which the sky position and distance are known exactly. 
We find that the \pz model performs as well as the method currently used within the LIGO-Virgo collaboration, but additionally it enables improving the measurement of specific components of the instrument control through astrophysical calibration.
\end{abstract}

\maketitle

\section{Introduction}
\label{sec:intro}

The advanced gravitational-wave (GW) detectors LIGO~\cite{Harry:2010zz,TheLIGOScientific:2014jea} and Virgo~\cite{TheVirgo:2014hva} have concluded their third observation run as of March 2020, reporting the detection of 56 candidate GW sources~\cite{GraceDB}, most of which, if confirmed, are binary black holes (BBHs).
Owing to planned increases in sensitivity for LIGO and Virgo, and the addition of the Japanese detector KAGRA~\cite{Akutsu:2018axf} to the global network, the detection rate will be even higher in the next few years~\cite{Aasi:2013wya}.
Having access to a large number of GW sources will allow for unprecedented measurements of the mass and spin distribution of compact objects, as well as their formation channels~\cite{LIGOScientific:2018jsj}. 
The potential of detecting many binary neutron star mergers (BNSs) together with electromagnetic (EM) counterparts opens the way to precise measurements of the Hubble constant~\cite{Schutz:1986gp,Monitor:2017mdv,Abbott:2017xzu,Chen:2017rfc,Abbott:2019yzh}.
Some of the detected sources will have high signal-to-noise ratio (SNR), which would enable precise tests of general relativity and of the nature of individual objects. 

For gravitational-wave astrophysics to fulfill its potential, one must control all of the (known) sources of systematics. 
In this work we focus on instrumental calibration uncertainties. 
The complex function that relates the voltage measured at the output of LIGO and Virgo photodetectors to the strain needed for astrophysical inference is the response function, $R(f)$. 
In the Fourier domain, the relation between these quantities is simply

\beq
d(f) \equiv \frac{\Delta L}{L} = R(f)\; v(f)
\label{Eq.data}
\eeq

where $v(f)$ is the photodetector readout, $d(f)$ is the gravitational-wave strain, $\Delta L$ is the differential arm (DARM) displacement of the mirrors, and $L$ is the nominal length of the interferometer arms~\cite{TheLIGOScientific:2016agk}. 
The calibration process includes collecting a set of measurements performed on the detectors to inform a reference model of \err~\cite{Abbott:2016jsd}, tracking the slow time-dependence of the detector response with respect to that model~\cite{Tuyenbayev:2016xey}, the use of that model to create a near-real-time data stream is an estimate of $d(f)$ at any time~\cite{Viets:2017yvy}, and characterizing the systematic error and statistical uncertainty in the model, or equivalently in the data stream used for astrophysical analysis~\cite{Sun:2020wke}.
The fundamental reference fiducials for the calibration process are independent laser systems, called \emph{photon calibrators} (Pcal),
 to calibrate LIGO and Virgo by applying a known radiation pressure directly into the test masses~\cite{Abbott:2016jsd,Karki:2016pht,Bhattacharjee:2020yxe}. 
Errors, bias, or uncertainty in any part of this calibration process to develop the estimated strain (including that in the Pcal systems) directly affect the strain, and hence
if unaccounted for, bias the estimation of the source parameters. 
Ref~\cite{Vitale:2011wu} has shown how the parameters that would suffer the largest biases are those mostly related to the amplitude of GW signals. 
For compact binaries coalescences (CBCs), those would be luminosity distance ($D_L$), orbital inclination ($\iota$), and sky position. 
In turn, those parameters are related to some of the key science goals mentioned above: identification of an EM counterpart and cosmology.

Statistical uncertainties and systematic errors in the measurement of the response function result in both amplitude and phase offsets, so that the actual response function at a specific time and frequency is related to the true response function by:

\beq
R(f,t) = \left(1+\delta A(f,t) \right) e^{i \delta \phi(f,t)} R^\mathrm{(true)}(f,t)
\label{Eq.RandRtrue}
\eeq

where \dea is the the \emph{relative} amplitude error and \dep the phase error. 
In turn, this affects the observed GW data as:

\beq
d_\mathrm{obs} = d_\mathrm{true} \left(1+\delta A(f,t) \right) e^{i \delta \phi(f,t)} 
\label{Eq.dobs}
\eeq

Here we are explicitly reporting a time dependence to stress that the behavior of GW detectors, and hence their transfer functions, varies over timescales of minutes~\cite{Tuyenbayev:2016xey}. 
Therefore, while it is generally a good approximation to treat the response function as constant in time (not in frequency) when analyzing a single CBC event, since its duration will usually be smaller than 2 minutes (for a BNS detected by advanced detectors), one should not assume that the response function is the same throughout an observing run. 
In fact, the response function of the LIGO and Virgo detectors is 
characterized continuously in a few small frequency bins throughout the run, and across all frequencies weekly, as a precaution against unexpected changes~\cite{2017PhRvD..96j2001C, Sun:2020wke}.

Currently, the results presented by the LIGO-Virgo collaboration (LVC) obtained with the \linf~\cite{Veitch:2014wba} or \texttt{Bilby}~\cite{2019ApJS..241...27A,Romero-Shaw:2020owr} source characterization algorithms marginalize over calibration errors 
with a spline interpolant informed by the frequency-dependent $68\%$ credible interval contours of the systematic error and uncertainty in each response function~\cite{FarrW} (henceforth \scal method).
While that approach has the advantage of accounting for calibration uncertainties, it also has some limitations. 
First, it introduces a significant number of nuisance parameters that must be marginalized over numerically: {roughly 20} parameters per interferometers. 
Second, these splines curves don't have any physical relation with the instrument, and do not enforce any physical correlation length (the frequencies at which the spline points are anchored are chosen uniformly in log space). 
Third, the spline marginalization method treats the uncertainties in the phase and amplitude of the response function as independent and uncorrelated. 
Fourth, should any constraints be placed on the response function through a so called astrophysical calibration (see below) it would be hard or impossible to related those constraints to specific components of the detector. 

In this paper we propose a new approach to account for uncertainties in the response function, which builds upon recent progress in measurement and modeling of the response function, and does not suffers from the same limitations of the spline approach.
We implement the new method, called ``physical calibration'' (Henceforth, \pz) in the \linf software and we apply it to the CBCs detected by LIGO and Virgo in their second observing run, as well as on simulated binary neutron star sources. 

The rest of this paper is organized as follows: in Sec.~\ref{SuSec.Calib} we summarize the measurements and algorithms used to calibrate the LIGO instruments; in Sec.~\ref{SuSec.PE} we present the implementation of the \pz method; in Sec.~\ref{SuSec.O2} and Sec.~\ref{SuSec.Injs} we report results from the analysis of LIGO-Virgo sources and simulated signals, respectively; finally in Sec.~\ref{sec:conclusions} we summarize the main conclusions.

\section{Method}
\label{sec:method}

\subsection{Calibration physical model}\label{SuSec.Calib}

While a full description of systematic error and uncertainty in the calibration of the LIGO detectors is beyond the scope of this paper, we will review the main points, and refer the interested reader to Ref.~\cite{Sun:2020wke} for more details.

In frequency domain, the complex-valued detector response can be written as
\begin{eqnarray}
\label{eqn:response}
R(f) = \frac{1}{C(f)} + A(f) D(f).
\end{eqnarray}
The sensing function $C$ converts the suppressed DARM residual displacement~\footnote{That is, the residual differential displacement of the mirrors after the control signal has been applied, see e.g. Fig. 3 of Ref.~\cite{Sun:2020wke}.} to digitized photo-detector output signals. 
The actuation function $A$ converts the requested digital control signal to the force applied to the test masses, producing a control displacement meant to suppress the DARM displacement.
The total $A$ function consists of three actuation stages, the upper intermediate (U), penultimate (P), and test mass (T) stages in the quadruple suspension~\cite{aston2012update}.
The function $D$ represents a set of digital, feedback control filters, which can be assumed as perfectly known. 
The DARM strain, and thus the calibrated data in Eq.~\ref{Eq.data}, are reconstructed using the modeled sensing and actuation functions, $C^{\rm (model)}$ and $A^{\rm (model)}$, in the detector calibration pipeline.
Here $A^{\rm (model)}$ denotes the model of the total $A$ function, in which each stage $A_a$ is modeled independently (\mbox{$a=U,P,T$}). 
The time-dependent, frequency-dependent systematic errors on our model of the response function
are written as
\begin{equation}
{\eta}_R = \frac{R^{\rm (true)}}{R^{\rm (model)}},
\label{eq:etaR}
\end{equation}
where $R^{\rm (true)}$ is the true detector response, and \mbox{$R^{\rm (model)}= 1/C^{\rm (model)} + A^{\rm (model)} D$} is the modeled response~\cite{Sun:2020wke}. 
The relative amplitude error and phase error in Eq.~\ref{Eq.dobs} can thus be written as
\begin{equation}
\begin{split}
\dea = & |{\eta}_R|-1, \\
\dep = & \angle {\eta}_R.
\end{split}
\label{eq:deltaAphi}
\end{equation}
Throughout the observing run, ${\eta}_R$ and its associated uncertainty is evaluated at a 1-hour cadence. 

The models $C^{\rm (model)}$ and $A^{\rm (model)}$ contain many parameters representing the entire DARM control loop from the basic properties of signal processing electronics to complex actuator dynamics. Most parameters can be measured independently to high precision, and do not dominantly contribute to the systematic error and/or uncertainty in $R^{\rm (model)}$.
However, a set of physical parameters related to specific properties of the instrument must be determined from interferometric measurements taken while the detectors are in the most sensitive, nominal operational state~\cite{Sun:2020wke}. 
These parameters, discussed as follows, highly depend on the loosely controlled alignment and thermal state of the detector and may vary slowly over time. They are difficult to measure and hence likely to introduce systematic errors in the calibration model.
For the sensing function, we write the physical parameter vector as
\begin{equation}
\bm{\lambda}^C = \left[H_C,  f_{cc},  f_s,  Q, \delta \tau_C\right],
\end{equation}
where $H_C$ is the overall gain of the sensing function, $f_{cc}$ is the differential coupled-cavity pole frequency, $f_s$ and $Q$ are, respectively, the pole frequency and quality factor of an optical spring-like response of any detuning between the coupled Fabry-P\'{e}rot arm cavities and signal recycling cavity~\cite{Hall2019}, and $\delta \tau_C$ is the residual time delay in $C$.
For the $a-$th stage of the actuation function ($a=U,P,T$), the physical parameter vector is
\begin{equation}
\bm{\lambda}^A_{a} = \left[H_a,  \delta \tau_a\right],
\end{equation}
where $H_a$ is the overall gain for the $a-$th stage actuator, and $\delta \tau_a$ is the residual time delay in that stage.
Some parameters in $C$ and $A$ vary slowly over time, on a time-scale of minutes to days, due to various physical mechanisms~\cite{TDCF-T1700106}. 
The overall gain variation of $H_C$ is tracked by a real-valued scalar factor $\kappa_C(t)$. 
Parameters $f_{cc}$,  $f_s$, and $Q$ in the sensing function are also time-varying. 
The variation of $H_a$ ($a=U,P,T$) is tracked by scalar factors $\kappa_U(t)$, $\kappa_P(t)$, and $\kappa_T(t)$ for each corresponding actuation stage. 
A full description of $C$ and $A$, as well as all the time-independent and time-dependent factors therein is given in Ref.~\cite{Sun:2020wke}. 

While $R^{\rm (model)}$ does an excellent job at reproducing $R^{\rm (true)}$, the residual systematic error ${\eta}_R$ and its uncertainty need to be quantified through the frequency-dependent, time-independent residuals $\eta_C = {C^{\rm (true)}}/{C^{\rm (model)}}$ and \mbox{$\eta_{A_a} = {A_a^{\rm (true)}}/{A_a^{\rm (model)}}$}, where $C^{\rm (true)}$ and $A^{\rm (true)}$ are the true sensing and actuation functions inferred from large collections of interferometric measurements, and the subscript $a$ indexes the actuation stages (\mbox{$a=U,P,T$}). 
This set of residuals is computed via Gaussian Process Regression (GPR)~\cite{GPR,scikit-learn}, using a physically motivated covariance kernel with model-agnostic hyperparameters that take into account potential frequency-dependent correlations. 
The posterior results from the GPR indicate the residual errors and uncertainties in the sensing and actuation models.
In a perfect calibration model, $\eta_C$, $\eta_{A_a}$, and hence $\eta_R$ are at unity in magnitude and zero in phase.

At any given time $t$, measurements of the various physical quantities that we have just described and that affect the response function (which we will collectively refer to as \mbox{\pz} parameters from now on) are used to assess the complex-valued, frequency-dependent systematic error in the detector response and its associated uncertainty. 
Using interferometric measurements, we apply Markov Chain Monte Carlo (MCMC) methods to obtain the posterior density functions of $\bm{\lambda}^C$ and $\bm{\lambda}^A$. 
The maximum a posteriori values of $\bm{\lambda}^C$ and $\bm{\lambda}^A$ are used to form the model functions $C^{\rm (model)}$ and $A^{\rm (model)}$, and thus $R^{\rm (model)}$. 
Since $\bm{\lambda}^C$ and $\bm{\lambda}^A$ are time-varying, the time-dependent corrections are taken into consideration when constructing $R^{\rm (model)}$ for any given analysis time. 
We use $10^4$ fair draws from the posterior probability density functions (PDFs) of $\bm{\lambda}^C$ and $\bm{\lambda}^A$ to create a distribution of draws from $R$ as described below. 
These $R$ samples, once divided by $R^{\rm (model)}$, yield a posterior distribution for ${\eta}_R(f;t)$.
The generic $i-$th sample for the inferred response function posterior reads~\cite{Sun:2020wke}

\begin{eqnarray}
{R_i} (f;t) &=&  \eta_{{\rm Pcal}_i}\left[\frac{1}{{\eta}_{C_i}(f) {C}(\bm{\lambda}_{i}^C;f;t)} \right.\nonumber\\
  &&+ {\eta}_{A_i}(f) {A}(\bm{\lambda}_{i}^A;f;t) {D}(f)\bigg].
\label{eq:rrnom}
\end{eqnarray}
The samples for the sensing and actuation functions ${C}(\bm{\lambda}_{i}^C;f;t)$ and ${A}(\bm{\lambda}_{i}^A;f;t)$ are derived from the MCMC posterior distributions of $\bm{\lambda}^C$ and $\bm{\lambda}^A$.
The samples ${\eta}_{C_i}(f)$ and ${\eta}_{A_i}(f)$ are, respectively, drawn from the GPR posterior distributions.
Here in Eq.~\ref{eq:rrnom}, we do not explicitly split out the three stages in $A$, and use ${\eta}_{A_i}$ to denote the sample of the residual in total $A$.
The $1\sigma$ uncertainties of the time-dependent factors applied in $C$ and $A$ at time $t$ are taken into account. 
The real-valued scale factor $\eta_{\rm Pcal}$ accounts for the systematic error and uncertainty of the photon calibrator, common to all interferometric measurements in a detector. 

The median frequency-dependent value of the $10^4$ samples from the distribution of ${\eta}_{R}(f;t)$ represents our best estimate for the systematic difference between $R^{\rm (true)}$ and ${R}^{\rm (model)}$ at time $t$, and thus the systematic error in the calibrated data $d(f;t)$.
Meanwhile, the 16th and 84th percentiles represent the bounds of systematic error and $1\sigma$ statistical uncertainty in the modeled detector response, and thus $d(f;t)$. 

For each of the LIGO detectors, we perform the above procedure and store to file the $10^4$ posterior samples from the posteriors of the \pz parameters, together with the resulting posterior samples for the frequency-dependent response function, Eq.~\ref{eq:rrnom}. 
Virgo detector does not have as sophisticated an infrastructure, but the detector response can be modeled in the same way~\cite{Virgo-Cal}.
The next section describes how these are used in the source characterization algorithm.

\subsection{Implementation in source characterization code}\label{SuSec.PE}

Given a stretch of interferometric data $d$ containing a CBC signal, one wants to estimate the posterior distributions of the unknown source parameters \vt (masses, spins, sky position, etc. See e.g. Ref~\cite{Veitch:2014wba}).
Bayes theorem allows to write:

\beq
p(\vt | d ) = \frac{p(d| \vt) \pi(\vt)}{p(d)}
\eeq

where $\pi(\vt)$ is the prior distribution of the CBC parameters (in what follow we will use the standard priors used by the LVC~\cite{LIGOScientific:2018mvr}) and $p(d)$ is the evidence of the data, which won't play a role in parameter estimation~\cite{Veitch:2014wba}. 
The remaining term is the likelihood of the data given \vt. 
If one assumes that the interferometric noise is stationary and Gaussian, then the likelihood in the Fourier domain reads:

\beq
p(d| \vt) \propto e^{- \langle d(f) - h(f,\vt) | d(f) - h(f,\vt) \rangle }
\label{Eq.Like}
\eeq

where we have defined the inner product:

\beq
\langle a | b \rangle \equiv 2 \int \ud f \frac{a b^* + a^* b}{S(f)}. 
 \eeq
 
 and $h(f,\vt)$ is the gravitational-wave template calculated at \vt. 
The likelihood weights the difference between data and GW template (i.e. the data residuals) by the noise power spectral density (PSD) $S(f)$~\cite{Sathyaprakash:2009xs,Chatziioannou:2019zvs}, i.e. the noise auto correlation. 
These expressions are written for a generic interferometer, and are extended to a network of detectors just by multiplying the likelihoods calculated in each interferometer~\cite{Veitch:2014wba}.

If one wants to explicitly account for statistical uncertainties and systematic errors in the response function, the likelihood in Eq.~\ref{Eq.Like} needs to be modified by correcting the data, Eq.~\ref{Eq.dobs}, or -- which is equivalent~\cite{FarrW} -- by modifying the GW template $h(f,\vt)$:

\beq
h(f,\vt) \rightarrow h(f,\vt) (1+\dea(\laa,\lac,f))e^{i \dep(\laa,\lac,f)}
\label{Eq.Template1}
\eeq

As mentioned in Sec.~\ref{SuSec.Calib}, the calibration pipelines produces draws from the posterior distribution of the response function errors, which can be used to obtain frequency-dependent medians and standard deviations for amplitude and phase errors, \dea and \dep. 
Current LVC results are produced by only using these medians and 1-sigma uncertainties to inform the position and width of the Gaussian priors of the calibration spline points~\cite{LIGOScientific:2018mvr, GWTC1_PE_release}.

Instead, we wish to augment \linf so that it can directly use \emph{individual} draws from \err, i.e. for \dea and \dep as defined by Eq.~\eqref{eq:deltaAphi}. 
We would also like to fold in the analysis the possibility that, beside interferometer-dependent amplitude errors, there may be a \emph{common} offset in the amplitude of the LIGO's response function introduced by the calibration of LIGO's Pcal lasers against a reference from the National Institute of Standards~\cite{Karki:2016pht}. 
We will use the variable \hpg to indicate this common offset.

We will thus work with the following template for the likelihood of LIGO's data:

\beqa
h^I(f,\vt)& \rightarrow &\hpg \;h^I(f,\vt) \left[1+\deaI(\laa,\lac,f)\right]\nonumber\\
&\times&e^{i \depI(\laa,\lac,f)}
\label{Eq.Template}
\eeqa

where and index $I=H$ ({Hanford}) or $L$ ({Livingston}) is used to label quantities which are instrument-dependent.

To run a source characterization analysis on a CBC event detected at some GPS time $t$ we thus proceed in two steps. 
First we build the distribution of frequency dependent systematic error, $\eta_{R}$, described in Sec~\ref{SuSec.Calib} for each of the LIGO detectors at time $t$. 
As described above, this produces a file with $10^4$ samples from the posteriors of the \pz parameters and their corresponding response function, which, given $R^{\rm{(true)}, I}$ at time $t$, can be recast into posteriors for \deaI and \depI following Eqs.~\eqref{eq:etaR} and~\eqref{eq:deltaAphi}.
We then deploy a modified version of \linf to generate posterior PDFs for both the CBC parameters \vt and the \pz parameters. 

More specifically, we modify the likelihood function, priors, and the sampler of \linf so that it can use the files containing $\eta_{R}^{I}$ and $R^{I, \rm{(model)}}$ directly. 
For each of the LIGO detectors:

\begin{itemize}
\item We load to memory the corresponding \pz file. We label each of the samples produced by the calibration pipeline with an integer from 1 to $10^4$;
\item We introduce a new sampling parameter, an integer between 1 and $10^4$, and assign it a uniform prior. We call it the \pz ID of this interferometer;
\end{itemize} 

The common \hpg parameter is assigned a uniform prior in the range $[-0.9914, 1.0086]$
consistent with the uncertainties on the calibration of the LIGO photon calibrators at the time of our analysis.
With these changes implemented, the parameter estimation algorithm proceeds as usual: at each iteration of the MCMC chain (or update of a Nested sampling live point~\cite{Skilling:2006gxv,Veitch:2014wba}), we update \vt, the calibration \pz IDs, \hpg, calculate the modified waveform templates for each interferometer, and hence the corresponding likelihood.
Our updates allow the user to use a different calibration marginalization scheme (\scal, \pz, no marginaization) for each detector independently, when running a network analysis. 
For the runs described in the remaining of this paper, we will only use the \pz method for the LIGO detectors, while using the spline method for Virgo.
In total, our scheme introduces a single new parameter for each instrument for which the \pz method is used, plus \hpg. 
This should be compared with the $\sim 20$ new parameters used for each instrument, if the spline method is used.

\section{Results}
\label{sec:results}

\subsection{Analysis of LIGO-Virgo's sources}\label{SuSec.O2}

In this section we apply the \pz method to all of the CBCs found by the LVC during their second observing run (O2), using the corresponding public data release~\cite{LIGOScientific:2018mvr, GWOSC_O2, GWTC1_PE_release}~\footnote{
We cannot re-analyze the sources detected in the first observing run, since the distribution of systematic error and uncertainty, $\eta_{R}$, was not recorded.}.
LIGO-only data is available for GW170104, GW170608 and GW170823, whereas LIGO-Virgo data is available for GW170729, GW170809, GW170814, GW170817 and GW170818. 

The Bayesian priors on the CBC parameters are chosen to match those used by LVC, whereas the priors on the \pz parameters have been described in the previous section.
We use the IMRPhenomPv2 waveform approximant~\cite{Hannam:2013oca, Husa:2015iqa, Khan:2015jqa} for all the BBH runs, with the reduced order quadrature (ROQ) likelihood implementation~\cite{Smith:2016qas}, while we use IMRPhenom\_NRTidal~\cite{Hannam:2013oca, Husa:2015iqa, Khan:2015jqa,Dietrich:2017aum, Dietrich:2018uni} for the binary neutron star binary GW170817~\footnote{The ROQ likelihood in \linf is distinct from the likelihood that is used for most waveform families. Our implementation of the \pz method works for both the ROQ and the ``classic'' likelihood.}.

\begin{figure}[hbt]
\includegraphics[width=0.45\textwidth]{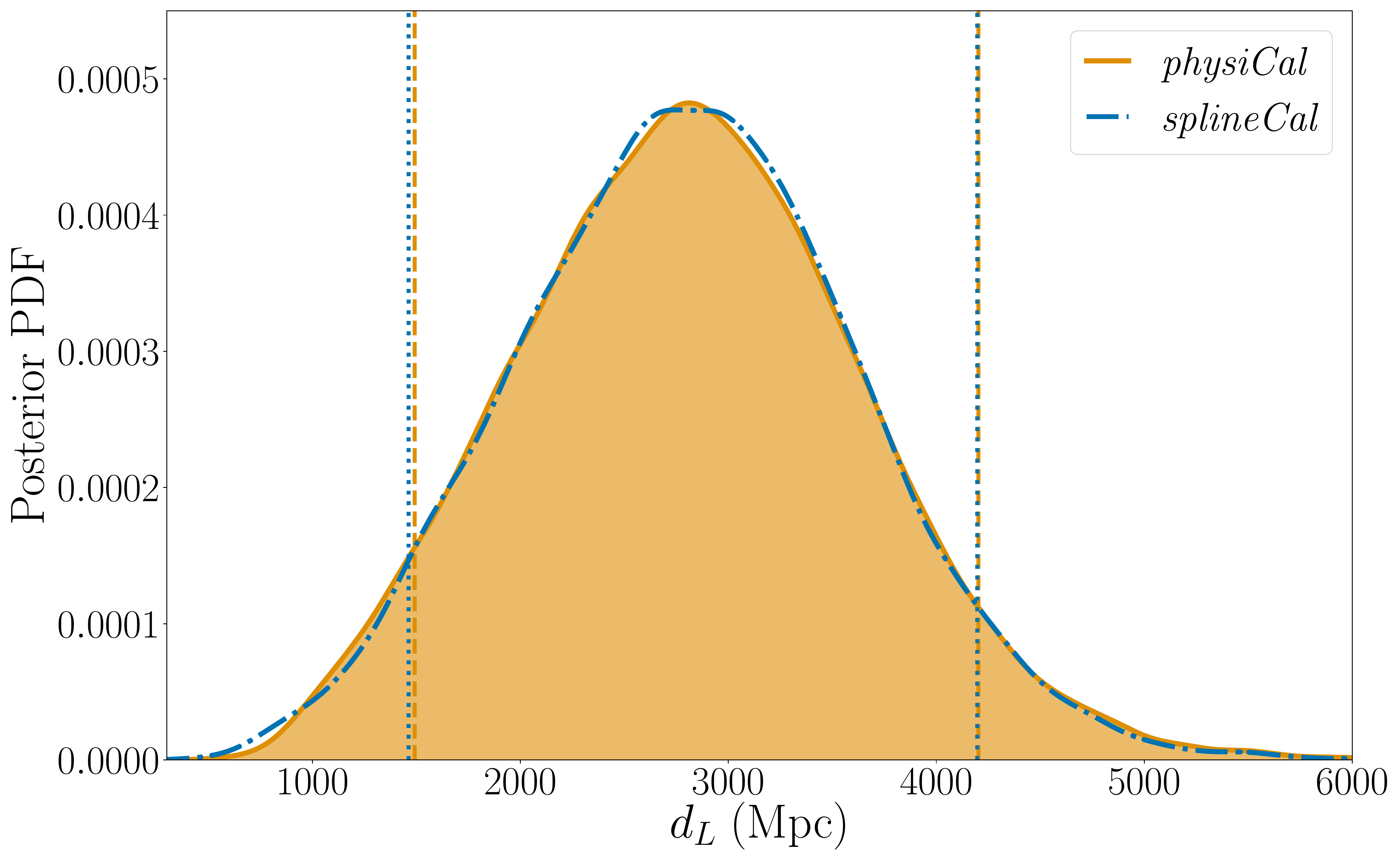}
\caption{Posterior density function for the luminosity distance of GW170729 inferred using the LVC's spline marginalization of the calibration uncertainty (\scal~\cite{FarrW,GWTC1_PE_release}) and the \pz method described in this work.
The vertical lines denote the 90\% credible interval for each analysis.}
\label{Fig. 170729_dL}
\end{figure}

For all sources, we find that the posterior distribution of the astrophysical parameters \vt obtained with the \pz method are virtually indistinguishable from those reported by the LVC using the spline method.
For example, in Fig.~\ref{Fig. 170729_dL} we show the posterior of the luminosity distance of GW170729, the most remote of the sources in the GWTC-1 catalog, obtained with \pz and with the spline method. 

{This can be explained by noticing that for both the spline and the \pz method no constraints can be placed on any of the calibration parameters, and the respective priors are recovered. Since the priors are informed by the same underlying calibration model, the two approaches yield consistent results. 
The O2 LVC sources~\cite{LIGOScientific:2018mvr} had network SNRs in the range $\sim[10,33]$. This suggests that even higher SNRs and/or some auxiliary information about the sources is needed to constrain the \pz parameters, Sec.~\ref{SuSec.Injs}.}
Ref.~\cite{Essick:2019dow} analyzed the BNS GW170817 with a different approach, and similarly found that nothing can be learned about the response function.

\begin{figure*}[hbt]
\includegraphics[width=0.45\textwidth]{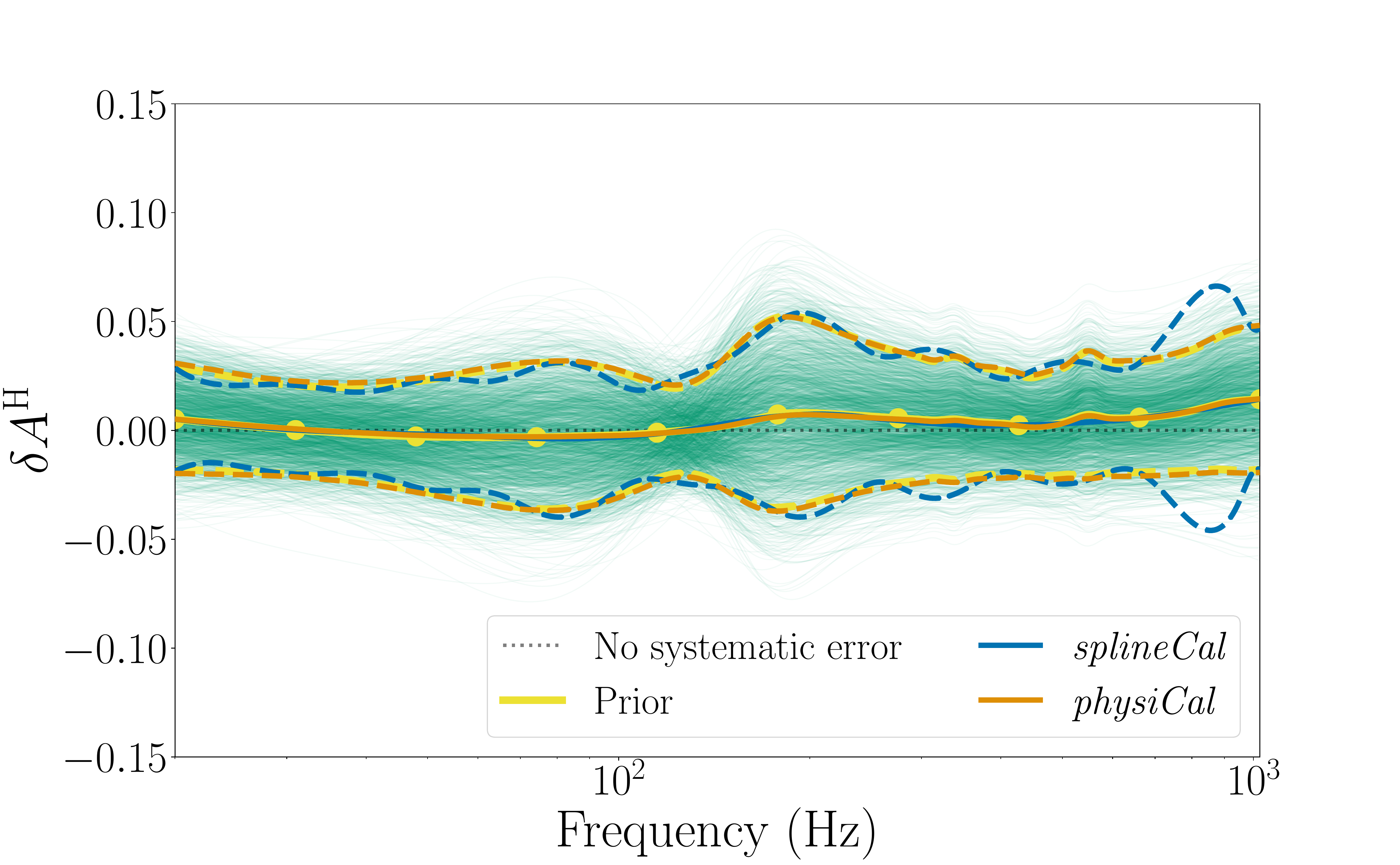}\includegraphics[width=0.45\textwidth]{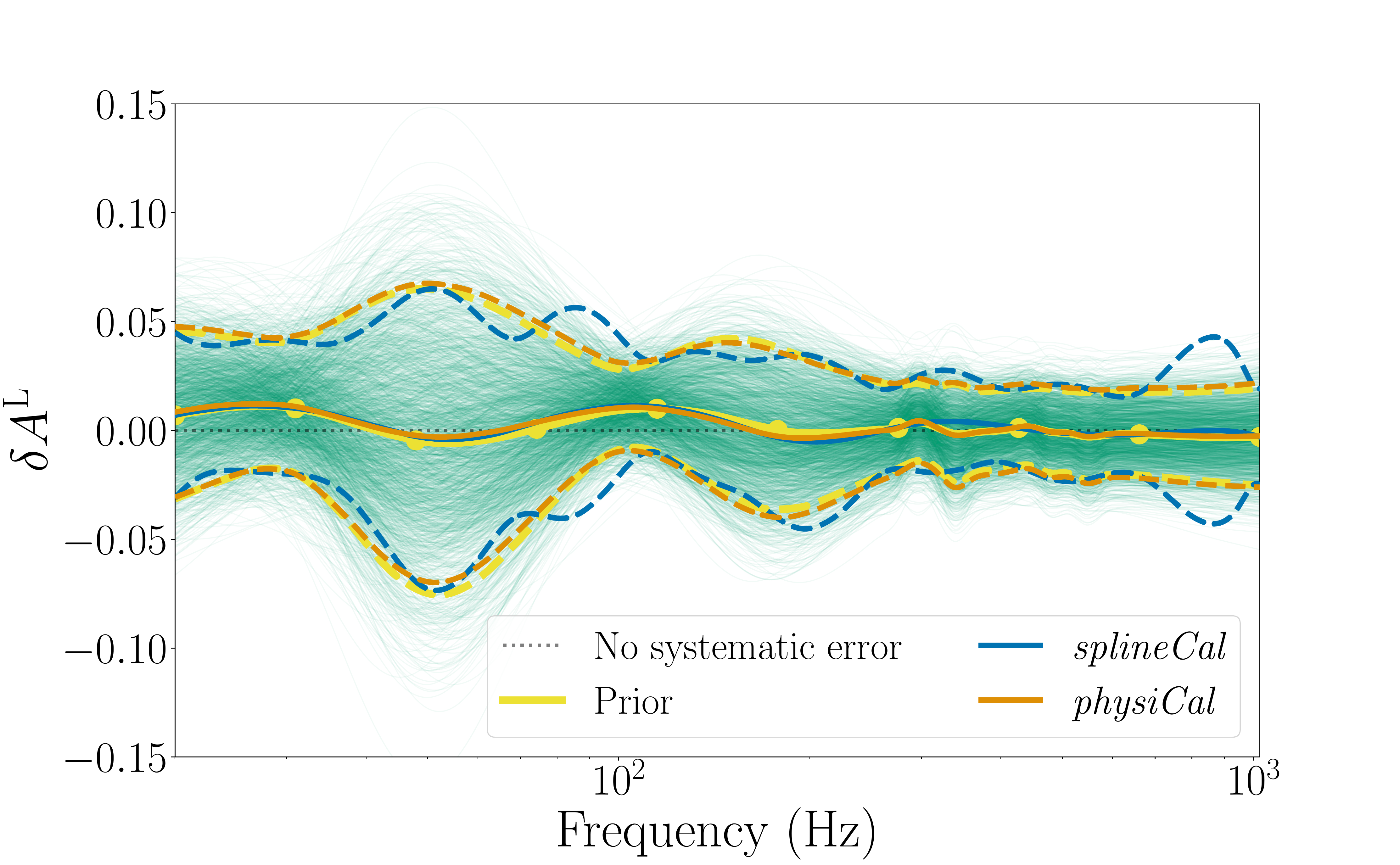} \\
\includegraphics[width=0.45\textwidth]{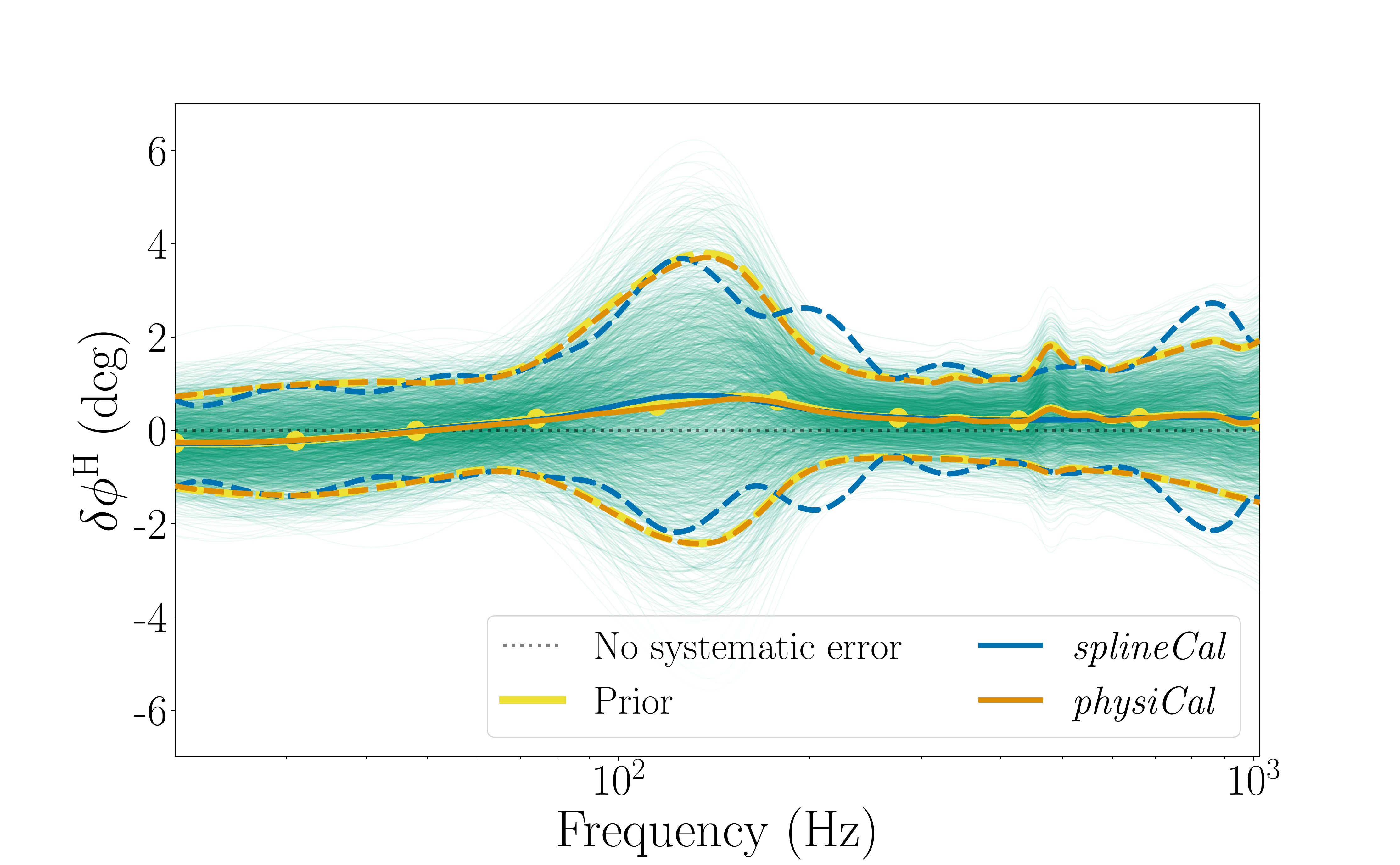}\includegraphics[width=0.45\textwidth]{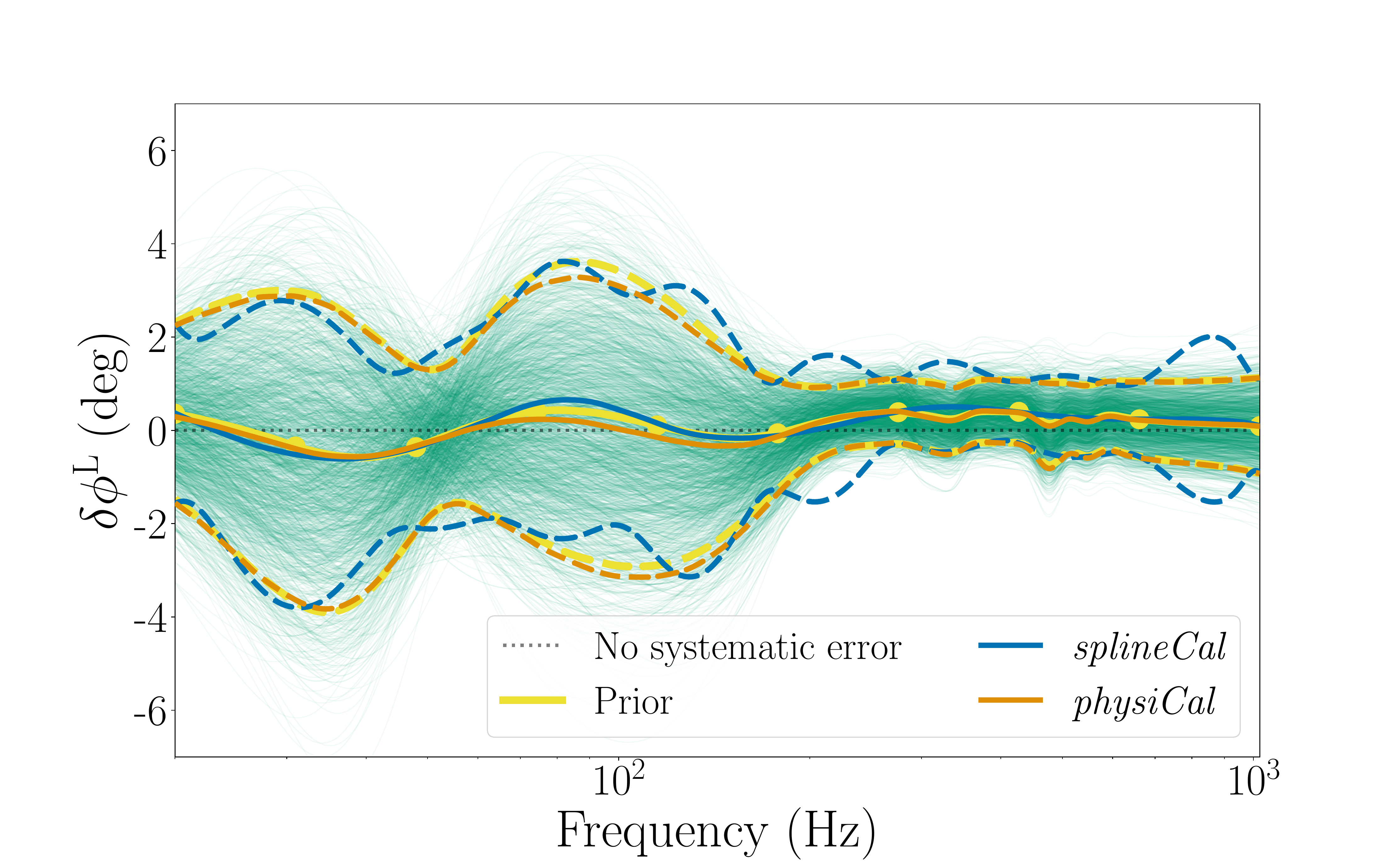} \\
\caption{ PDFs for the amplitude (top) and phase (bottom) of the response function's errors at the time of GW170814 for the LIGO Hanford (left) and LIGO Livingston (right) detectors. 
The grey dotted line indicate the ideal case with no systematic error.
All PDFs are represented by their median (solid line) and $90\%$ credible interval (dashed lines)
The prior distributions are shown in yellow.
The yellow dots indicate the frequencies where the \scal variables are defined.
The \scal posteriors are shown in blue and the \pz posteriors are shown in orange.
For the \pz method, we also show 2048 individual draws from the posteriors (green semi transparent curves).}
\label{Fig.calib_compare}
\end{figure*}

In Fig.~\ref{Fig.calib_compare} we show a comparison of the posteriors for the response function's errors when analyzing {GW170814}. 
Amplitude and phase errors are reported -- for the two LIGO detectors individually -- in the top and bottom rows respectively. 
The blue lines refer to the spline method and the orange lines to the \pz method. 
In both cases, the solid lines are the medians and the dashed lines mark the 90\% credible intervals. 
For the \pz method, we also show 2048 individual draws from the posteriors (semi transparent green curves).

\subsection{Simulated events}\label{SuSec.Injs}

The results we obtained for the O2 sources show that with the ``typical'' CBC source of medium-low SNR for which most or all of the astrophysical parameters are unknown, no information can be gained about the \pz parameters, and we just recover the priors. 
This might be due to the fact that the effect of calibration errors mostly affects the amplitude of the response function, and hence of the signal~\cite{Vitale:2011wu}. On the other hand, analysis of CBC signals cannot usually constrain amplitude parameters as well as parameters that affect the phase evolution of the system (e.g. chirp mass)~\cite{LIGOScientific:2018mvr}.
Since our ignorance of the calibration parameters would result on a broadening of the signal's amplitude which is much smaller than the uncertainty due to other factors e.g. distance-inclination degeneracy~\cite{TheLIGOScientific:2016wfe,Chen:2018omi,Usman:2018imj}, little or no information can be usually gained about the \pz parameters. The situation could be radically different if the the model for the response function were significantly off, which might be visible in the posteriors of the \pz parameters. Multiple factors could be result in large and unaccounted for residual errors in the response function: an imperfect evaluation of time-dependent terms, 

On the other hand, if extra astrophysical information is obtained about the signal that decreases the correlations between CBC parameters, one might hope to set constraints on the \pz parameters. 
While the idea of ``astrophysical calibration'', i.e. of learning something about the detector using particularly loud or otherwise exceptional events is not new~\cite{Pitkin:2015kgm,Essick:2019dow}, we stress that the best one can do using the spline approach is to verify that something is wrong with the overall response function.
With the \pz method instead, we can hope to say something about specific parts of the sensing and actuation systems, as described in Sec.~\ref{SuSec.Calib} above and Refs.~\cite{Viets:2017yvy,2017PhRvD..96j2001C,Sun:2020wke}.

To verify if this is the case, we add 200 simulated BNSs into \emph{real} LIGO-Virgo interferometric data from O2~\cite{GWOSC_O2} (we only consider BNSs and not BBHs because will want to assume the source extrinsic parameters can be constrained with EM data, see below).
The signals' merger times are randomly chosen to be in the 3600 seconds preceding or following the 8 CBC sources detected in LIGO-Virgo's second observing run~\footnote{If the simulated signal precedes the O2 detection, we leave enough time between them to avoid overlaps.}. 
Rather than producing the full distribution of $\eta_{R}$ for each simulated event, we re-use the distributions at the time of the 8 O2 sources.
For each of the simulated signals we thus use the output of the calibration pipeline as calculated for the nearest of the O2 sources.
This implies that the largest possible time interval between the time a simulated signal is added into the data and the assigned O2 event time for which its $\eta_{R}$ was produced is one hour.
The simulated events are assigned random sky positions and orbital orientation, and are placed uniformly in comoving volume. 
This implies that the resulting signal-to-noise ratios (SNRs) are representative of realistic detections in the second and third observing runs (i.e. with network SNRs in the approximate range $[10,40]$ and with most sources having SNR near the minimum).
\begin{table*}[t]
\centering
\caption{The true values of some selected parameters for the two BNS sources described in Sec.~\ref{SuSec.Injs}. \label{Tab.Injs}
}
\begin{tabular}{!{\vrule width \heavyrulewidth}c!{\vrule width \heavyrulewidth}c!{\vrule width \heavyrulewidth}c!{\vrule width \heavyrulewidth}c!{\vrule width \heavyrulewidth}c!{\vrule width \heavyrulewidth}c!{\vrule width \heavyrulewidth}c!{\vrule width \heavyrulewidth}c!{\vrule width \heavyrulewidth}c!{\vrule width \heavyrulewidth}} 
\hline
\hline
ID & $m_1 \;[M_\odot]$ & $ m_2\; [M_\odot]$ & $D_L \;[\mathrm{Mpc}]$ & $t$           & $\iota\; [\mathrm{rad}]$ & SNR$_H$ & SNR$_L$ & SNR$_V$  \\ 
\hline
1 &1.98            & 1.78             & 58.8        & 1167560557.32 & 0.22          & 21.4    & 21.9    & n/a      \\ 
\toprule
2& 1.99            & 1.69             & 74.4        & 1187057243.40 & 0.71          & 13.5    & 25.9    & 3.3      \\
\hline
\hline
\end{tabular}
\end{table*}
For these analyses, we use the IMRPhenomPv2 waveforms both to simulate the signals that are added into the data, and for the parameter estimation algorithm. 
The neutron stars are assigned randomly oriented spins with (dimensionless) magnitude uniform in the range $[0,0.2]$ and component masses uniform in the range $[1.8-2.4] M_\odot$~\footnote{This range of mass was not chosen to be representative of a realistic mass distribution, but rather to optimize the runtime of \linf with the ROQ likelihood}. 
We do not include tidal effects either in simulating signals nor in the subsequent source characterization analysis

To mimic a situation where a successful electromagnetic counterpart has been found, which yields the source's 3D position, we run the source characterization algorithm by assuming that the sky position \emph{and} the luminosity distance of the sources are perfectly known. This neglects potential uncertainties introduced by the cosmology used to convert the source redshift into a luminosity distance; however, here we are interested into a somewhat optimistic scenario to show what this method can theoretically do. If, as it is more realistic, the distance to the source is only known within an uncertain range, the overall amplitude parameter \hpg would not be constrained. 
While it is possible to also obtain some constraints about the source orbital inclination by folding in external information about the source~\cite{Mandel:2017fwk,Finstad:2018wid}, that inference would not be very precise and would depend on detailed modeling of the EM emission. 
Therefore, instead of assuming the inclination angle is perfectly known, we restrict its Bayesian prior to a $\pm 20^\circ$ interval symmetric around the true value excluding unphysical values (i.e. $\iota<0$ rads and $\iota>\pi$).
Having fixed luminosity distance and sky position to their true values, the inclination angle is thus the only CBC parameter that significantly affects the amplitude of the signals in our analysis~\footnote{Intrinsic parameters also affect the GW amplitude. However they are usually measured from the GW phase well enough that they can be thought as known when considering the signal's amplitude.}.
It is worth stressing that even for LIGO-only analyses, \hpg is not perfectly degenerate with the (cosine of the) inclination angle, since this latter affects the two GW polarizations each in a different way~\cite{Sathyaprakash:2009xs}, while the former is an overall amplitude offset.
This would be different if the luminosity distance were also a free parameter, since in that case \hpg and distance would be perfectly degenerate in a LIGO-only analysis, and only the combination $\hpg/D_L$ would be measurable.

We will not report extensively on these simulations because for the overwhelming majority of them, owing to the low SNRs, nothing is learned about the \pz parameters. 
Instead, we will just focus on two high-SNR signals, one in HVL data, and the other in HL data. 
The true values of some of their parameters are reported in Tab.~\ref{Tab.Injs}, together with the ID we will use to refer to each.

\begin{figure*}[tb]
\includegraphics[width=0.45\textwidth]{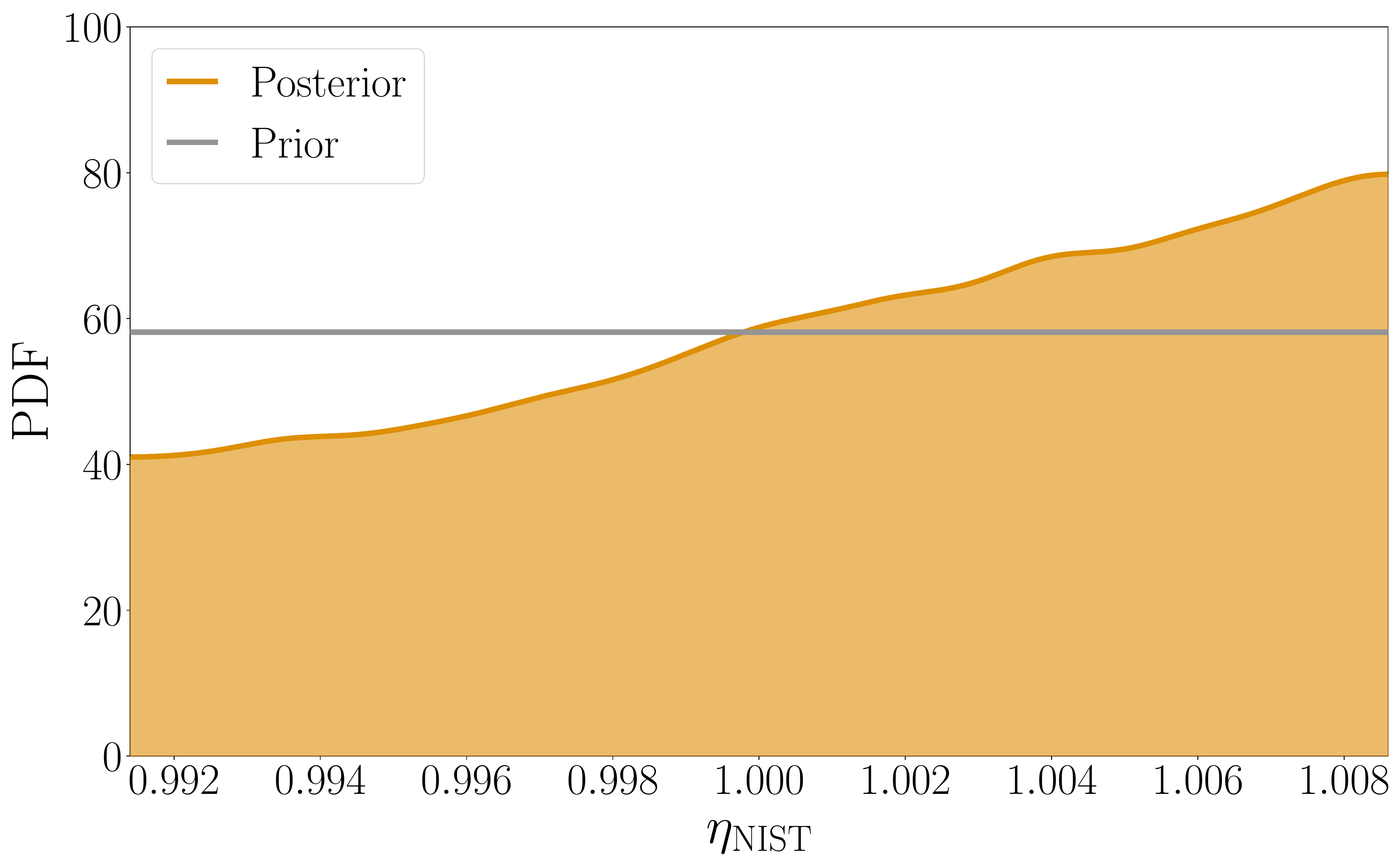}\includegraphics[width=0.45\textwidth]{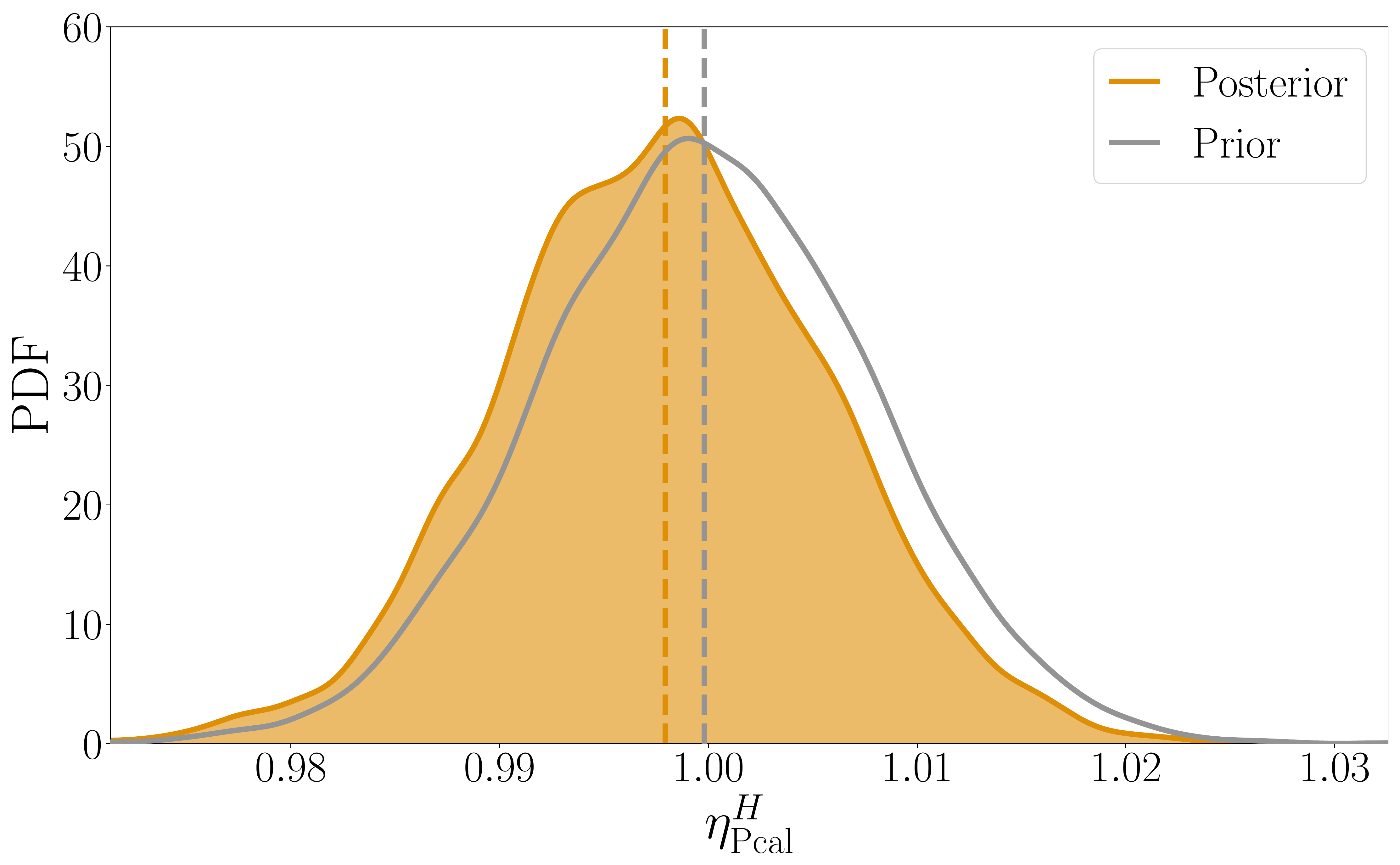}\\
\includegraphics[width=0.45\textwidth]{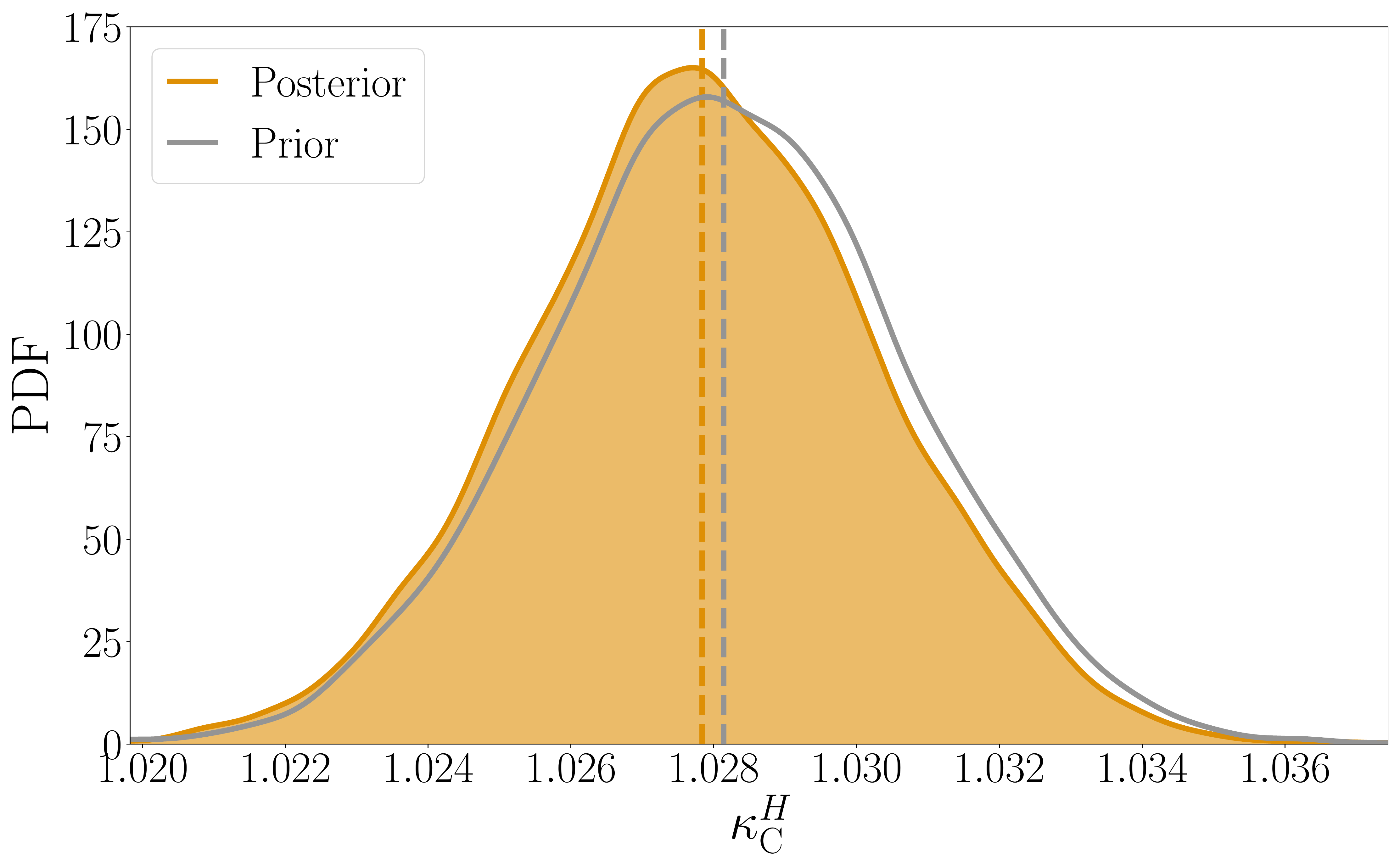}\includegraphics[width=0.45\textwidth]{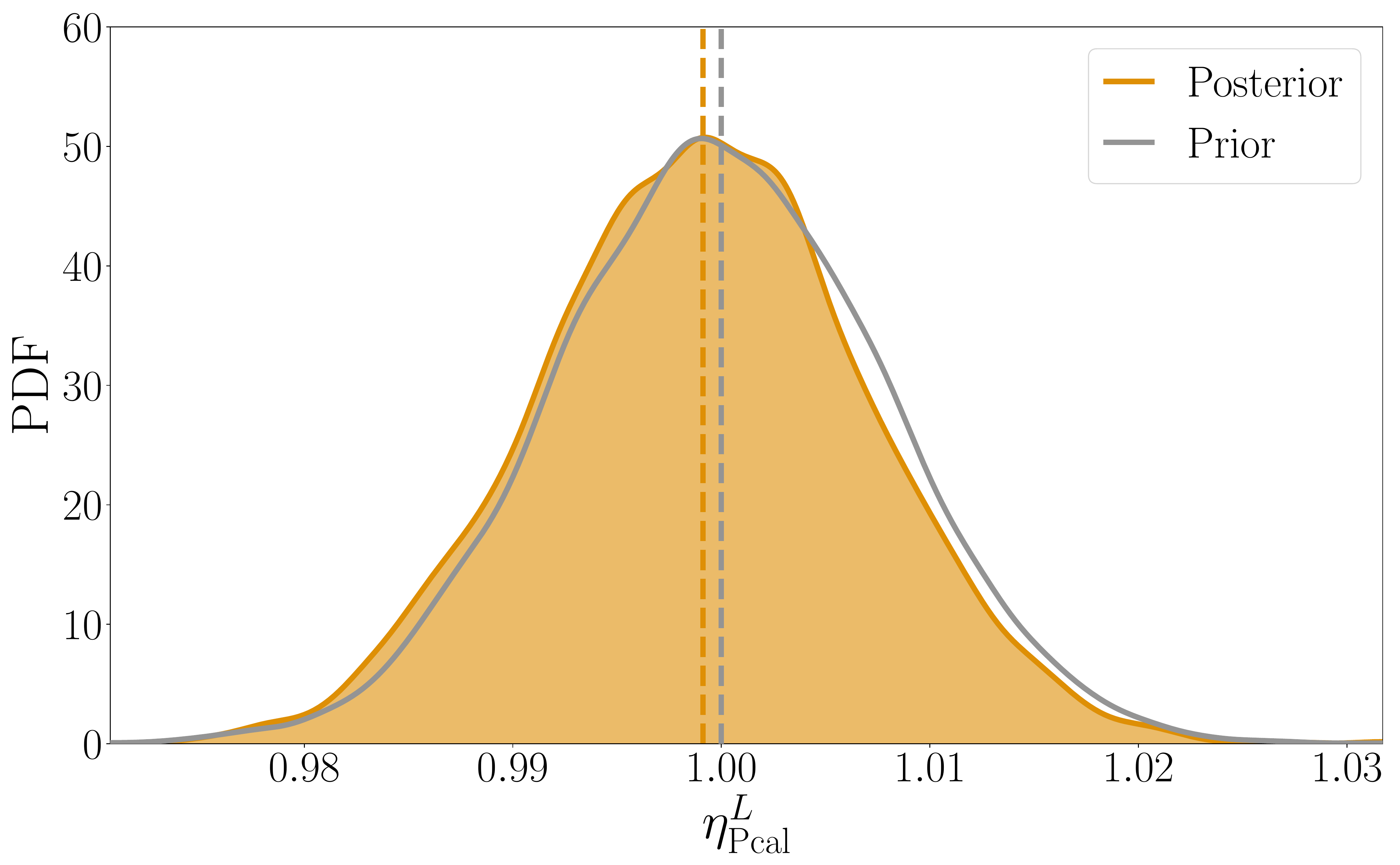}
\caption{Posterior distributions for the \pz parameters for which information is gained relative to the priors, for the \mbox{BNS \#1}, (see Tab.~\ref{Tab.Injs}). 
The respective priors are shown as solid grey lines.
The median of each PDF is shown as a dashed vertical line.}\label{Fig.Ev1Post}

\end{figure*}

The BNS \#1 is added into LIGO-only data, since Virgo was not operating at the time. 
While for most of the \pz parameters the prior is returned, a handful of posterior distributions are informative and are shown in Fig~\ref{Fig.Ev1Post}, together with their priors.
We see that the posterior of \hpg, while still broad and with support in the whole prior range, does have some some support for values larger than one. 
Meanwhile, the posterior for $\eta_{{\rm Pcal}}^H$, which controls the overall amplitude of the response function in LIGO Hanford, is clearly different from its Gaussian prior, and prefers slightly smaller values. 
For $\eta_{{\rm Pcal}}^L$, the corresponding parameter for LIGO Livingston, the effect is not as significant.
The other parameter that shows a slight departure from its priors is $\kappa_C^H$, a time-dependent parameter related to the sensing function of LIGO Hanford~\cite{Sun:2020wke}. 
We quantify the similarity between the prior and posterior distributions through a Jensen-Shannon (JS) divergence statistic~\cite{Lin_ShannonEntropy}, a general symmetrized extension of the Kullback-Leibler divergence~\cite{kullback1951}.
The JS-divergence is defined over the range 0 bits (identical distributions) to 1 bit (no statistical overlap).
For the \pz parameters shown in Fig~\ref{Fig.Ev1Post}, the JS-divergences are 0.11 bits ($\eta_{{\rm Pcal}}^H$), 0.09 bits (\hpg), 0.05 bits ($\kappa_C^H$) and 0.05 bits ($\eta_{{\rm Pcal}}^L$) respectively.
In all these cases, we see that the offsets are much smaller than the statistical uncertainties. 
The posteriors of all other \pz parameters are either even more similar to, or undistinguishable from, their priors.

\begin{figure}[htb]
\includegraphics[width=0.45\textwidth]{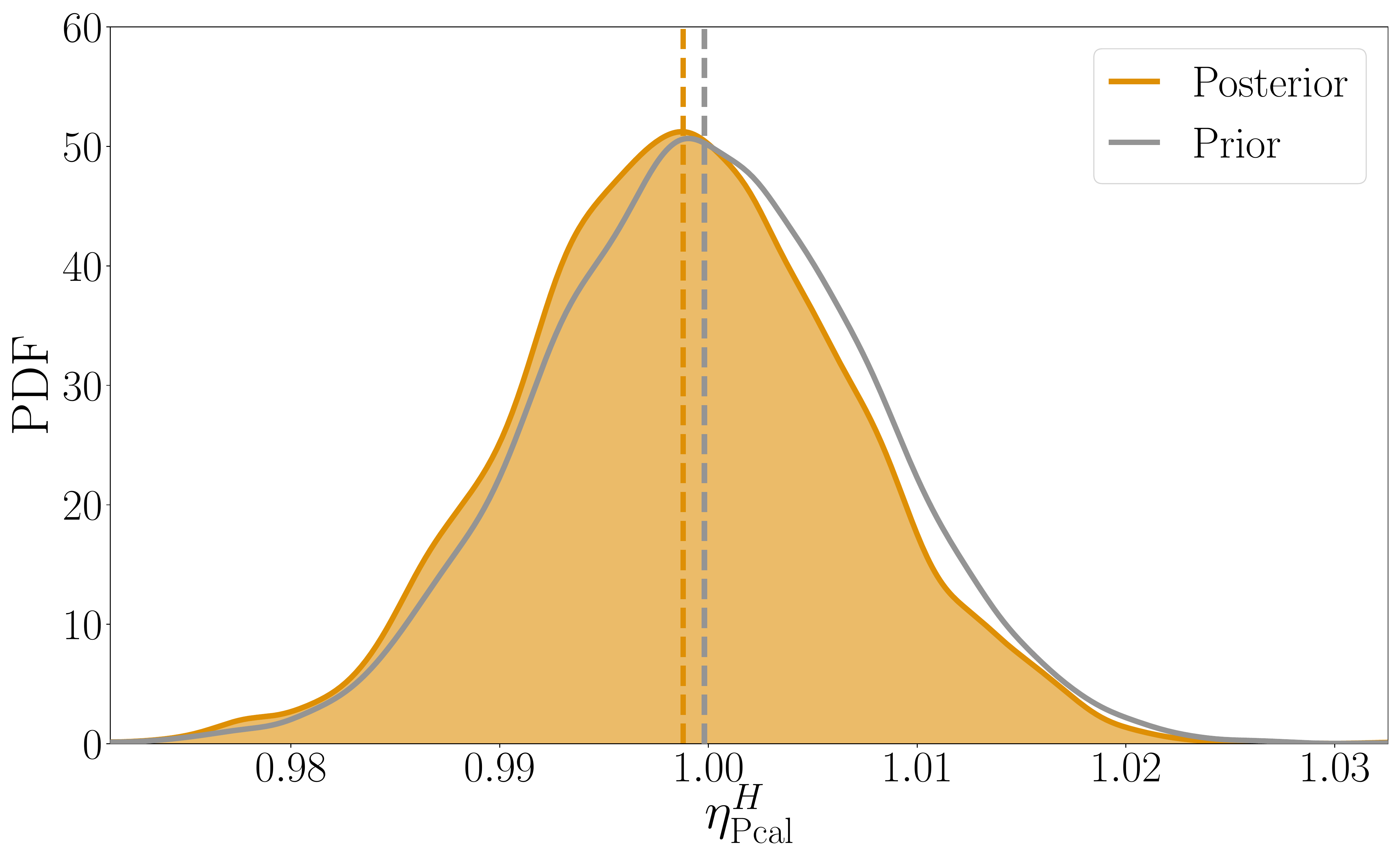}
\caption{Posterior distribution of $\eta_{{\rm Pcal}}^H$ for the BNS \#2 (see Tab.~\ref{Tab.Injs}). 
The prior is shown as a solid grey line.
The median of each PDF is shown as a dashed vertical line.}\label{Fig.HpcalEv2}
\end{figure}

When considering the BNS \#2 we find instead that {all} of the \pz parameters return exactly the prior, except $\eta_{{\rm Pcal}}^H$ (shown in Fig.~\ref{Fig.HpcalEv2}), for which the JS-divergence is 0.06 bits. 
Thus, despite a comparable network SNR and the presence of Virgo, less information is gained about the \pz parameters for the BNS \#2 than for the BNS \#1.
This suggests that the SNR is not the only figure of merit to predict if and what can be learned with astrophysical calibration. 
Instead, this might be suggestive of the fact that the \emph{model} for the response function was adequate for the BNS \#2, whereas it was not for the BNS \#1. 
Theoretically, it is possible that even if the model for the response function is correct, we could beat the statistical uncertainties on the \pz parameters, i.e. obtain posteriors which are centered at the same positions as their priors, but are narrower. 
We speculate that a similar measurement would require even higher SNRs, and will explore that possibility in a future publication.

\section{Conclusions and outlook}
\label{sec:conclusions}

In this paper we have proposed a different and more physical approach to marginalizing over possible systematic error associated with the calibration of ground-based gravitational-wave detectors, called \pz. 
We account for departures from the nominal value of the instruments' response function using directly the output of the calibration pipeline of the LIGO's instruments (the method can be extended to other detectors, even though we have not done it for this study). {This method improves the existing approach, which relies on unmodeled spline curves to model calibration errors, hence discarding some of the available information about the detectors and their response functions.}

We have augmented the \linf source characterization algorithm with the \pz method, and used it to analyze the 8 CBC signals in the public data from the second observing run of the LVC. 
We find that the posteriors for the CBC parameters obtained with \pz are extremely similar to those produced by the LVC with the existing spline method. 
This should not be too surprising since either method is not constraining at the SNRs that can be expected given the current detectors' sensitivities, and de facto produces posteriors for the calibration parameter (spine points or \pz parameters) equal to their priors. 
We then looked at the possibility of astrophysical calibration, i.e. he idea that an high SNR CBC observation, with perfectly known extrinsic parameters derived from an accompanying electromagnetic characterization, can be used to learn something about systematic error in each detector's calibration.
We created a set of simulated BNS signals and added them to real public data from the LVC's second observing run. 
For all analyses, we assumed that the sources' sky positions and luminosity distances are perfectly known, whereas the orbital inclination angles are known to within $20^\circ$. 
This is meant to mimic a very successful EM campaign which provides information about position and orientation of the binaries.
We find that for most of the simulations nothing can be learned about the \pz parameters, and the posteriors are very similar to their priors. 
Only with the loudest BNSs we considered, with network SNRs around 30, we obtained posteriors for some of the \pz parameters that were clearly, though not dramatically, different from their priors. 
Furthermore, we found that the SNR is not the only relevant parameter to forecast how informative any given source will be, and we showed that two BNSs with virtually the same SNRs can yield quite different posteriors for the \pz parameters.
Ultimately, both a high SNR and an imperfect model for the response function at the time of the simulated event are necessary to see variations. 
In the representative system we showed, the parameters that were most different from their priors were the overall amplitude and two of the parameters associated with the sensing function in the LIGO Hanford detector. 
To us, this is one of the main advantages of the \pz method over the spline method: that astrophysical calibration can potentially yield information about \emph{specific} components involved with the calibration process, rather than about the response function as a whole.
While for some of the loudest BNSs we considered we observed some departure from the modeled response function, the uncertainty in the \pz parameters was not narrower than the prior uncertainty established by the calibration pipeline. 
That is, some of the posteriors shifted relative to their priors, but maintained the same shape. 
It is possible that with even louder signals one could decrease the prior statistical uncertainty in the \pz parameters. {A large scale study will be necessary to explore more systematically the parameter space, and fully understand which sources would yield the best astrophysical calibration, and which of the \pz parameters are more likely to be constrained.}
Another possible avenue to improve our understanding of the response function, is combining multiple detections. 
In fact, even though for most of the weaker sources very little is learned about the instrument, one can potentially combine all detected signals and build joint posteriors for the subset of the \pz parameters that do not depend on time, and are thus expected to have the same value throughout a science run. 
Both of these prospects will be explored in a future publication.

\emph{Note:} After this work had begun, an independent group started exploring the possibility of using importance sampling to marginalize over physical calibration parameters~\cite{Ethan}. The two methods yield consistent results.

\section{Acknowledgements}
We thank Paul Lasky, Ethan Payne, Colm Talbot and Eric Thrane for useful discussion and for sharing an early version of their manuscript.\\
SV, CJH, LS and JK acknowledge support of the National Science Foundation and the LIGO Laboratory.
EG acknowledges the support of the Natural Sciences and Engineering Research Council (NSERC) of Canada.
LIGO was constructed by the California Institute of Technology and Massachusetts Institute of Technology with funding from the United States National Science Foundation, and operates under cooperative agreement PHY--1764464.
Advanced LIGO was built under award PHY-- 0823459. 
The authors are grateful for computational resources provided by the LIGO Laboratory and supported by National Science Foundation Grants PHY-- 0757058 and PHY-- 0823459.
This research has made use of data, software and/or web tools obtained from the Gravitational Wave Open Science Center (https://www.gw-openscience.org), a service of LIGO Laboratory, the LIGO Scientific Collaboration and the Virgo Collaboration.
This analysis was made possible by the {\tt LALSuite}~\cite{lalsuite}, {\tt numpy}~\cite{numpy}, {\tt SciPy}~\cite{Virtanen:2019joe} and {\tt matplotlib}~\cite{Hunter:2007ouj} software packages. 
The authors would like to thank all of the essential workers who put their health at risk during the COVID-19 pandemic, without whom we would not have been able to complete this work.
This is LIGO Document Number DCC-P2000293.

\appendix

\bibliographystyle{apsrev4-1}
\bibliography{draft}

\end{document}